\documentclass[submission,Phys]{SciPost}
\usepackage{graphicx,bm}
\usepackage{amsmath}
\usepackage{amsfonts}
\usepackage{amssymb}
\usepackage{braket}
\usepackage{dsfont}
\usepackage{bm}
\usepackage{eucal}
\usepackage{mathtools}
\usepackage{comment}

\usepackage{bbm}

\usepackage{color}

\usepackage{xcolor}

\usepackage{tikz}
\usetikzlibrary{calc}
\usetikzlibrary{decorations.markings}
\usetikzlibrary{decorations.pathmorphing, arrows}
\usetikzlibrary{arrows.meta}

\newcommand{\be}{\begin{equation}}
\newcommand{\ee}{\end{equation}}
\newcommand{\bea}{\begin{eqnarray}}
\newcommand{\eea}{\end{eqnarray}}
\newcommand{\up}{\uparrow}
\newcommand{\down}{\downarrow}

\def\sfix#1{\texorpdfstring{#1}{Lg}}
\def\doi{http://dx.doi.org/}
\def\nn{\nonumber\\}

\def\fr#1{(\ref{#1})}

\def\sgn{\mathrm{sgn}}
\def\eps{\epsilon}

\def\bol{\boldsymbol{\lambda}}
\def\bon{\boldsymbol{\nu}}
\def\d0{\{d,d,0\}}

\begin{document}
\begin{center}
    \Large{\bf Interacting hydrodynamic modes in spinless fermions with dephasing noise}
\end{center}
	\date{\today}
	\begin{center}
	Fabian H.L.\ Essler and Patrik Penc, 
\end{center}	
\begin{center}
The Rudolf Peierls Centre for Theoretical Physics, University of Oxford, OX1 3PU, UK\\
\end{center}
\section*{Abstract}
{\bf 
We study the non-equilibrium dynamics of spinless fermions with dephasing noise in the framework of a many-particle Lindblad equation. Using a mapping to the exactly solvable one-dimensional Hubbard model with purely imaginary tunneling amplitude we analyze the Heisenberg-picture dynamics of operators quartic in fermions and determine their hydrodynamic projections. We construct the relevant diffusive eigenoperators explicitly and show that, in the quartic sector, they can be interpreted as interacting pairs of bilinear hydrodynamic modes. As a consequence, translationally invariant quartic operators generically exhibit non-vanishing diffusive late-time tails, unlike translationally invariant bilinears. Our results show that the hydrodynamic tails of microscopic operators cannot in general be inferred from symmetry constraints or coarse-grained fluctuating hydrodynamics alone; they also depend crucially on the spatial structure and effective size of the hydrodynamic eigenoperators.
}

\date{\today}
\tableofcontents

\section{Introduction}
\label{sec:intro}
Dissipative dynamics in many-particle quantum systems has attracted a great deal of attention in recent years \cite{fazio2025many}. Of particular interest have been situations in which the environmental degrees of freedom can be treated in a Markovian approximation, which can be described in terms of Lindblad equations(LEs)\cite{gorini1976completely,lindblad1976generators,breuer2002theory}. These are generally still difficult to analyze, which motivated the search for LEs that can be mapped to interacting Yang-Baxter integrable
models~\cite{Medvedyeva2016Exact,ziolkowska2020yang,Rowlands2018noisy,Naoyuki2019dissipative,nakagawa2021exact,buca2020bethe,Ribeiro2019integrable,essler2020integrability,robertson2021exact,yuan2021solving,de2021constructing,ishiyama2025exact,ishiyama2025bexact}.
Importantly, the corresponding Lindbladians are not quadratic in bosons or fermions, and do not preserve Gaussianity of density matrices under time evolution. The paradigmatic example is the tight-binding model of spinless fermions with dephasing noise \cite{Medvedyeva2016Exact},
\begin{align}
    \partial_t \rho(t) = \mathcal{L}[\rho(t)] = -i[H, \ \rho(t)] + \gamma \sum_{j=1}^L\big( 2n_{j}\rho(t)n_{j}-\{n_j,\ \rho(t)\}\big) ,
\end{align}
where $\gamma>0$ is the dephasing rate and the Hamiltonian on a ring of $L$ sites reads
\begin{equation}
    H = -J \sum_{j=1}^L c^\dagger_j c_{j+1} + c^\dagger_{j+1} c_{j}\ ,\quad
    \{c_j,c^\dagger_\ell\}=\delta_{j,\ell}\ .
\end{equation}
Throughout this work we set $J\geq 0$.
It was shown in Ref.~\cite{Medvedyeva2016Exact} that vectorizing the Lindblad equation results in an imaginary-time Schr\"odinger equation for the density matrix, where the associated  non-Hermitian Hamiltonian can be mapped onto the one-dimensional Hubbard model with purely imaginary hopping amplitude. The latter is solvable by Bethe ansatz \cite{essler2005one}, which allows for 
the exact determination of spectral properties of the Lindbladian \cite{Medvedyeva2016Exact}. Several recent works \cite{bhat2025transfer,ishiyama2025exact,ishiyama2025bexact,penc2026linear} went on to derive exact results for certain dynamical properties. In particular, in \cite{penc2026linear} we obtained exact results for the diffusive late-time Heisenberg-picture evolution of operators quadratic in fermions. This allowed us to obtain exact results for hydrodynamic late-time tails of local operators in diffusive systems. In particular we showed that
\begin{align}
\big(c^\dagger_{x_0}c_{y_0}\big)(t\gg\gamma^{-1})&\simeq
\sum_{x,y} \psi_{x_0,y_0}(x,y;t) c^\dagger_xc_y\ ,
\end{align}
where the amplitudes $\psi_{x_0,y_0}(x,y;t)$ are
expressed in terms of confluent hypergeometric functions.
At asymptotically late times they reduce to
(decaying) power-laws in $t$ \cite{bhat2025transfer}, with exponents that depend on 
$X = |x-y|+|x_0-y_0|$ and $Y=(x-x_0)+(y-y_0)$
\begin{equation}
\psi_{x_0,y_0}(x,y;t)\Bigg|_{\rm leading}=
\begin{cases}
A_e\ (\mathcal{D} t)^{-\frac{1+X}{2}} & \text{Y even}\ ,\\
A_o\ (\mathcal{D} t)^{-\frac{2+X}{2}} & \text{Y odd}\ .
\end{cases}
\label{2tails}
\end{equation}
Here the diffusion constant $\mathcal{D}$ is
\be
\mathcal{D}=\frac{J^2}{\gamma}\ .
\ee
We further showed that translationally invariant operators bilinear in fermions do not exhibit hydrodynamic late-time tails in the large volume limit
\be
(\sum_x c^\dagger_xc_{x+d})(t\gg\gamma^{-1})\simeq\mathcal{O}(e^{-2\gamma t}).
\ee
We argued that the origin of this behaviour are kinematic constraints that arise in the equations of motion for fermion bilinears in the class of Lindblad equations with decoupled BBGKY hierarchies \cite{haken1973exactly,Esposito_2005,eisler2011crossover,vzunkovivc2014closed,caspar2016dissipative,Caspar2016dynamics,Hebenstreit2017Solvable,Foss-Feig2017solvable,klich2019closed,PhysRevB.102.100301,lychkovskiy2021closed,bhat2025transfer,teretenkov2026fermionic}. We conjectured that translationally invariant charge-neutral operators quartic and higher in fermions do exhibit diffusive late-time tails.

The key objectives of this manuscript are as follows. 
We first present a physical explanation for the complicated structure of hydrodynamic tails in the tight-binding model with dephasing noise and other models with decoupled BBGKY hierarchies \fr{2tails}. We explain that the structure of the tails is mainly determined by the spatial sizes of the hydrodynamic eigenoperators of the Lindbladian, while bounds imposed by symmetry constraints are very loose. We believe that this could be an important, general principle that applies to other models as well. In particular it might explain some of the findings for diffusive Floquet dynamics in Ref.~\cite{matthies2026thermalization}. 
We then turn to the explicit construction of diffusive eigenoperators quartic in fermions. We determine their explicit forms and show how they are related to diffusive eigenoperators quadratic in fermions. In a superoperator formalism this relation takes the form of interactions between diffusive eigenmodes. 
Our main technical result is an explicit hydrodynamic projection formula for translationally invariant quartic operators. For density-density operators, the leading contribution decays as $t^{-1/2}$, in contrast to translationally invariant bilinears, whose hydrodynamic projections vanish in the thermodynamic limit. This difference is traced to the structure of the quartic diffusive eigenoperators, which are scattering states of two hydrodynamic bound states. Our analysis makes use of the integrability of the model \cite{Medvedyeva2016Exact} in an essential manner. 

\section{Lindblad equation and vectorization}
\label{sec:Lindblad}

The time evolution of Heisenberg-picture operators takes a simple form \cite{Medvedyeva2016Exact,penc2026linear}
\begin{align}
\frac{\mathrm{d}\mathcal{O}(t)}{\mathrm{d}t}= i[H, \ \mathcal{O}(t)] - \gamma \sum_{j=1}^L [n_{j},\ [n_{j},\ \mathcal{O}(t)]]\equiv \mathcal{L}^*[\mathcal{O}(t)] .
\label{Leqn}
\end{align}
In the following we will be interested in eigenoperators $\hat\Phi_\alpha$ of the dual Lindbladian
\be
\mathcal{L}^*[\hat\Phi_\alpha]=E_\alpha\hat\Phi_\alpha\ ,
\label{eigenops}
\ee
with eigenvalues such that ${\rm Re}(E_\alpha)\approx 0$. 
It is useful to vectorize the Lindblad equation \fr{Leqn} following \cite{penc2026linear}
\begin{align}
\mathds{1}_j&\longrightarrow |0\rangle_j\ ,\quad
c^\dagger_j\longrightarrow c^\dagger_{j,\uparrow}|0\rangle_j\ ,\quad
c_j\longrightarrow (-1)^jc^\dagger_{j,\downarrow}|0\rangle_j\ ,\quad
c^\dagger_jc_j\longrightarrow (-1)^jc^\dagger_{j,\uparrow} c^\dagger_{j,\downarrow}|0\rangle_j\ .
\label{vectorization}
\end{align}
This turns \fr{Leqn} into an imaginary-time Schr\"odinger equation
\be
\frac{\mathrm{d}}{\mathrm{d}t}|\mathcal{O}(t)\rangle=\mathcal{H}|\mathcal{O}(t)\rangle
\ee
with a non-Hermitian Hamiltonian
\be
{\cal
  H}=-iJ\sum_{j,\sigma}\big(c^\dagger_{j,\sigma}c_{j+1,\sigma}+{\rm
  h.c.}\big)+2\gamma\sum_j\big[
  (n_{j,\uparrow}-\frac{1}{2}) (n_{j,\downarrow}-\frac{1}{2})-\frac{1}{4}\big].
\label{HHub}
\ee
This is related to the usual Hubbard Hamiltonian
\be
H(U)=-\sum_{j,\sigma}\big(c^\dagger_{j,\sigma}c_{j+1,\sigma}+{\rm
  h.c.}\big)+U\sum_j\big[
  (n_{j,\uparrow}-\frac{1}{2}) (n_{j,\downarrow}-\frac{1}{2})\big]
\ee
by
\be
{\cal H}=iJ H(U=-\frac{2i\gamma}{J})-\frac{\gamma L}{2}\ .
\label{hubbardrel}
\ee
The eigenoperators \fr{eigenops} of the dual Lindbladian map onto the eigenstates of the Hubbard-model \fr{HHub}, which can be derived by a nested Bethe-Ansatz \cite{lieb1968absence,deguchi2000thermodynamics}. 
\subsection{Symmetries of the dual Lindbladian}
\label{ssec:symmetries}
The Hubbard model possesses an SO(4) symmetry with generators \cite{essler2005one}
\begin{align}
S^+&=\sum_{j=1}^Lc^\dagger_{j,\up}c_{j,\down}=(S^-)^\dagger\ ,\quad S^z=\frac{1}{2}\sum_{j=1}^L (n_{j,\up}-n_{j,\down})\ ,\nn
\eta^\dagger&=\sum_{j=1}^L(-1)^jc^\dagger_{j,\down}c^\dagger_{j,\up}\ ,\quad \eta^z=\frac{1}{2}\sum_{j=1}^L (n_{j,\up}+n_{j,\down}-1)\ .
\end{align}
These commute with the vectorized form of the dual Lindbladian 
\be
[\mathcal{H},S^\alpha]=0=[\mathcal{H},\eta^\alpha]\ ,\quad \alpha=\pm,z.
\ee
This implies that eigenstates of $\mathcal{H}$ can be labelled by the
numbers $N_{\uparrow,\down}$ of spin-up and spin-down
fermions. Moreover, given a right eigenstate of $\mathcal{H}$ one can
construct new eigenstates by acting with the SO(4) raising and
lowering operators \cite{essler1992new,essler1991complete}. The sector
$N_\up=N_\down=1$ was analyzed in detail in our recent work
\cite{penc2026linear}. In the following we extend this analysis to the
sector $N_\up=N_\down=2$.
The dual Lindbladian also has discrete symmetries. In the vectorized
picture they take the following forms:
\begin{itemize}
\item{} Translational invariance $T\mathcal{H}T^\dagger=\mathcal{H}$, where
\be
T c_{j,\sigma}T^\dagger=c_{j+1,\sigma}\ .
\ee
\item{} Parity invariance $P\mathcal{H}P=\mathcal{H}$, where
\be
P c_{j,\sigma}P=c_{L+1-j,\sigma}\ .
\ee
  
\end{itemize}

\section{Time evolution of operators}
\label{sec:timeevol}
Our objective is to determine the Heisenberg-picture time evolution of neutral
operators  
\begin{equation}
\mathcal{O}^{(2n)}(t=0) = \sum_{\boldsymbol{x}}
\psi_\mathcal{O}^{(2n)}(\boldsymbol{x})\prod_{j=1}^n
c^\dagger_{x_j}\prod_{k=n+1}^{2n}c_{x_k}\ .
\end{equation}
As a result of the decoupling of the BBGKY hierarchy, the time evolved
operators can be expressed as \cite{penc2026linear} 
\begin{equation}
\mathcal{O}^{(2n)}(t) = \sum_{\boldsymbol{x}}
\psi_\mathcal{O}^{(2n)}(\boldsymbol{x};t)\prod_{j=1}^n
c^\dagger_{x_j}\prod_{k=n+1}^{2n}c_{x_k}\ .
\end{equation}
In order to determine the late-time behaviour of the amplitudes
$\psi_\mathcal{O}(\boldsymbol{x};t)$ we proceed as follows: 
\begin{enumerate}
\item We vectorize the operator of interest
    \begin{equation}
        \mathcal{O}^{(2n)}\longrightarrow|\mathcal{O}^{(2n)} \rangle \ .
    \end{equation}
\item{} We determine the un-normalized left and right eigenstates of the Hamiltonian
$\mathcal{H}$
\begin{align}
|\Phi_{R,\alpha}\rangle&=\sum_{\boldsymbol{x}}\Phi_{R,\alpha}(\boldsymbol{x})\prod_{j=1}^nc^\dagger_{x_j,\down}
\prod_{\ell=n+1}^{2n}c^\dagger_{x_\ell,\up}|0\rangle\ ,\nn
\langle\Phi_{L,\alpha}|&=\sum_{\boldsymbol{x}}\Phi^*_{L,\alpha}(\boldsymbol{x})\langle
0|\prod_{\ell=2n}^{n+1}c_{x_\ell,\up}\prod_{j=n}^1c_{x_j,\down}\ ,
\end{align}
where $\boldsymbol{x}=(x_1,\dots,x_{2n})$.
These states fulfil
\begin{align}
\mathcal{H}|\Phi_{R,\alpha}\rangle&=E_\alpha|\Phi_{R,\alpha}\rangle\ ,\quad
\langle\Phi_{L,\alpha}|\mathcal{H}=E_\alpha\langle\Phi_{L,\alpha}|\ ,\nn
\langle\Phi_{L,\alpha}|\Phi_{R,\alpha'}\rangle&=A_{\alpha}^{-2}\delta_{\alpha,\alpha'}\ ,\quad
\mathds{1}=\sum_{\alpha}A_{\alpha}^2|\Phi_{R,\alpha}\rangle\langle\Phi_{L,\alpha}|\ ,
\label{LR}
\end{align}
and are simultaneous eigenstates of the translation operator
\be
T|\Phi_{R,\alpha}\rangle=e^{iP_\alpha}|\Phi_{R,\alpha}\rangle\ ,
\ee
where $P_\alpha$ are the momentum eigenvalues. We focus on eigenstates for which the corresponding eigenvalues have real parts close to zero. These dominate the late-time behaviour and are characterised by $\alpha$ such that
\be
\mathcal{S}=\{\alpha|\text{Re}\big(E_\alpha\big)>-\delta\}\ ,
\ee
where $\delta>0$ is a cutoff.
\item We expand the time-evolved vectorized operator in the eigenbasis of the Hamiltonian
    \begin{equation}
        |\mathcal{O}(t) \rangle =  \sum_{\alpha}e^{E_\alpha t}|A_{\alpha}|^{2}\langle\Phi_{L,\alpha}|\mathcal{O} \rangle |\Phi_{R,\alpha}\rangle\ .
    \end{equation}
At late times only eigenstates with ${\rm Re}(E_\alpha)\approx 0$ contribute
\begin{equation}
|\mathcal{O}(t\gg \gamma^{-1}) \rangle \approx \sum_{\alpha\in \mathcal{S}}e^{E(\alpha)t}|A_{\alpha}|^{2}\langle\Phi_{L,\alpha}|\mathcal{O} \rangle |\Phi_{R,\alpha}\rangle\ .
\label{specrep}
\end{equation}
In the $L\to\infty$ limit we turn sums into integrals and derive closed-form asymptotic expansions for the late-time form of Heisenberg-picture operators $\mathcal{O}(t)$.
\end{enumerate}
\subsection{Translationally invariant operators}
We are particularly interested in translationally invariant operators of the form
\be
\hat{\Psi}_{\boldsymbol{d}}=\sum_{j=1}^L c^\dagger_jc^\dagger_{j+d_1}c_{j+d_2}c_{j+d_3}.
\label{psid}
\ee
These describe the expectation values after quantum quenches from translationally invariant initial states $\rho(0)$
\be
{\rm Tr}\big[\rho(t)c^\dagger_jc^\dagger_{j+d_1}c_{j+d_2}c_{j+d_3}\big]=
\frac{1}{L}{\rm Tr}\big[\rho(t)\hat{\Psi}_{\boldsymbol{d}}\big].
\ee
An important result of our recent work \cite{penc2026linear} was that translationally invariant operators quadratic in fermions have vanishing hydrodynamic projections. We argued that this is a consequence of kinematic constraints that arise for quadratic operators as a result of the decoupling of the BBGKY hierarchy, and that operators quartic, sextic, etc in fermions would in general have non-vanishing diffusive tails. In the following we prove this conjecture by deriving explicit late-time expressions for Heisenberg-picture operators, which have the form
\be
\hat{\Psi}_{\boldsymbol{d}}(t)=t^{-\alpha_{\boldsymbol{d}}}\hat{\Psi}^{(1)}_{\boldsymbol{d}}+\text{ subleading}.
\ee
For translationally invariant operators the spectral representation
simplifies as we only need to take into account zero momentum
eigenstates of $\mathcal{H}$ in \fr{specrep},
i.e. replace $\mathcal{S}$ by
\be
\mathcal{S}_T=\{\alpha|P_\alpha=0 \text{ and } \text{Re}\big(E_\alpha\big)>-\delta\}.
\label{ST}
\ee

\section{What determines the hydrodynamic tails of microscopic operators?}
\label{sec:tails}
In view of heuristic approaches to hydrodynamic tails such as
\cite{matthies2026thermalization} it is important to establish the physical principles that determine them. As we will now show there are two key aspects: (i) symmetry constraints and (ii) the size and structure of the hydrodynamic eigenoperators. While the role of (i) was stressed in Ref.~\cite{matthies2026thermalization}, implications of the (ii) have to the best of our knowledge not been explored in the literature.

To understand their respective roles it is sufficient to consider the quadratic operators
\be
\hat{\mathcal{O}}_{m,\pm}=c^\dagger_{\frac{L}{2}-m}c_{\frac{L}{2}+m+1}\pm c^\dagger_{\frac{L}{2}+m+1}c_{\frac{L}{2}-m}\ .
\ee
Assuming that $L/2$ is an even integer their vectorized expressions are
\be
|{\cal O}_{m,\pm}\rangle=(-1)^{m+1}\big(c^\dagger_{\frac{L}{2}-m,\up}c^\dagger_{\frac{L}{2}+m+1,\down}\mp c^\dagger_{\frac{L}{2}+m+1,\up}c^\dagger_{\frac{L}{2}-m,\down}\big)|0\rangle.
\ee
The time-evolved operators take the form
\be
\hat{\mathcal{O}}_{m,\pm}(t)=\sum_{x,y}\psi_{m,\pm}(x,y;t)c^\dagger_x c_y\ ,
\ee
where the late-time asymptotic behaviour of the wave function is given by
\begin{align}
&\psi_{m,\pm}(x,y;t) \simeq \sum_{\alpha\in\mathcal{S}} \langle\Phi_{L,\alpha}|{\cal O}_{m,\pm}\rangle e^{E_\alpha t} \Phi_{R,\alpha}(y,x)\ (-1)^{y+1}.
\label{asympt}
\end{align}
At late times the wave-functions decay as power laws that are determined by the momentum dependence of the overlaps $\langle\Phi_{L,\alpha}|{\cal O}_{m,\pm}\rangle$ and the wave-functions $\Phi_{R,\alpha}(x,y)$ describing the diffusive eigenmodes \cite{penc2026linear}. This can be understood by noting that the diffusive modes correspond to $\alpha$ such that $P_\alpha\approx\pi$
\be
E_\alpha=-\frac{J^2}{\gamma}(P_\alpha-\pi)^2+\dots
\ee
\subsection{Symmetry constraints give bounds on hydrodynamic tails}
Symmetry constraints arise from the overlaps $\langle\Phi_{L,\alpha}|{\cal O}_{m,\pm}\rangle$. To show this we note that
the parity operation acts on the two states as
\be
\mathcal{P}|{\cal O}_\pm\rangle=\mp|{\cal O}_\pm\rangle\ ,\qquad
\langle\Phi_{L,\alpha}|\mathcal{P}=e^{-iP_\alpha}\langle\Phi_{L,\bar{\alpha}}|\ ,
\ee
where $P_{\bar{\alpha}}=-P_\alpha$ and the state $\langle\Phi_{L,\bar{\alpha}}|$ is obtained from $\langle\Phi_{L,{\alpha}}|$ by reversing the sign of all rapidity variables in the Bethe Ansatz wave functions. This leads to the following relation for matrix elements
\begin{align}
  \langle\Phi_{L,\alpha}  |{\cal  O}_{m,\pm}\rangle&=
\langle\Phi_{L,\alpha}|\mathcal{P}\mathcal{P}|{\cal
  O}_{m,\pm}\rangle=\mp e^{-iP_\alpha}\langle\Phi_{L,\bar{\alpha}}|{\cal O}_{m,\pm}\rangle\ .
\label{symmarg}
\end{align}
The diffusive eigenmodes occur around $P_\alpha\approx\pi$ \footnote{We recall that this is a consequence of the staggered factor $(-1)^j$ in \fr{vectorization}, which ensures that the hopping term in the Hamiltonians $\mathcal{H}$ have standard form.} and the matrix elements then have expansions of the form
\be
\langle\Phi_{L,\alpha}  |{\cal  O}_{m,\pm}\rangle\simeq b^{(0)}_{m,\pm}+b^{(1)}_{m,\pm}(P_\alpha-\pi)+\dots
\ee
The symmetry relation \fr{symmarg} implies that
\be
b^{(0)}_{m,-}=0\ .
\label{symmconst}
\ee
The late-time asymptotics of \fr{asympt} in the large-$L$ limit is obtained by turning the sum into an integral over the total momentum $P$ \cite{penc2026linear} and \fr{symmconst} then implies symmetry constraints on the leading asymptotic behaviour
\be
\psi_{m,\pm}(x,y;t)\propto (\mathcal{D}t)^{-\alpha_{m,\pm}}\ ,\quad \alpha_{m,+}\geq\frac{1}{2}\ ,\ \alpha_{m,-}\geq 1\ .
\label{symmbound}
\ee
\subsection{Role of the size and structure of the hydrodynamic eigenmodes}
The amplitudes $\Phi_{R,\alpha}(x,y)$ encode the spatial structure of the diffusive eigenoperator of the dual Lindbladian. For the class of models considered in \cite{penc2026linear} it has the structure of a two-particle bound state, the size of which is comparable to the lattice spacing and depends on the total momentum. More precisely we have
\be
\Phi_{R,\alpha}(x,y)\propto e^{iP_\alpha\frac{x+y}{2}}\left[\frac{iJ}{2\gamma}|\pi-P_\alpha|\right]^{|x-y|}\ ,
\label{wavefuncontr}
\ee
which implies that the effective size of the diffusive eigenoperator with total momentum $P$ is
\be
\ell(P\approx\pi)=\frac{1}{\ln\big(\frac{2\gamma}{J|\pi-P|}\big)}.
\ee
Importantly this depends on $P$, is not large compared to the lattice spacing, and vanishes when $P\to\pi$, which corresponds to on-site $\eta$-pairs
\be
\lim_{P\to 0}|\Phi_{R,\alpha}\rangle\propto\eta^\dagger|0\rangle\ .
\ee
The fact that $\Phi_{R,\alpha}(x,y)$ with $P\approx \pi$ has a non-trivial structure on the lattice scale also affects the overlaps
\begin{align}
\langle\Phi_{L,\alpha}  |{\cal  O}_{m,\sigma}\rangle&=(-1)^{m}\left[\Phi_{L,\alpha}^*(\frac{L}{2}+m+1,\frac{L}{2}-m)
-\sigma\Phi_{L,\alpha}^*(\frac{L}{2}-m,\frac{L}{2}+m+1)\right]\nn
&\propto\delta_{\sigma,-}\left[\frac{iJ}{2\gamma}|\pi-P_\alpha|\right]^{2m+1}\ .
\label{overlapcontr}
\end{align}
The combination of the momentum dependencies from \fr{overlapcontr} and \fr{wavefuncontr} gives the hydrodynamic tails  \cite{penc2026linear}
\be
\psi_{m,\sigma}(x,y;t)\propto \delta_{\sigma,-}(\mathcal{D}t)^{-m-\frac{3}{2}}\ .
\ee
These decay much faster than the bounds \fr{symmbound} arising from symmetry considerations. The latter are therefore ineffective with regards to predicting hydrodynamic tails of "microscopic" operators which act non-trivially only on a finite number of neighbouring sites in the thermodynamic limit. We note that the situation would be very different if the effective size of the diffusive eigenoperators were approximately constant and large compared to the lattice spacing
\be
\lim_{P\to \pi}\ell(P)=\ell_0\gg 1.
\ee
In such a scenario we would expect the hydrodynamic tails to follow the symmetry bounds as long as $m\ll\ell_0$. The general lesson of the above considerations is that power-law hydrodynamic tails of microscopic operators cannot in general be inferred from symmetry considerations, but can depend strongly on the spatial structure of the hydrodynamic eigenoperators. If the latter are characterized by a finite length scale $\ell_0$, symmetry arguments can be sufficient to infer hydrodynamic tails of microscopic operators with a spatial extent small compared to $\ell_0$. Alternatively one can consider appropriately coarse-grained observables that transform nicely under symmetries.
The above discussion has focused on operators quadratic in fermions, but we will see in the following that the same conclusions hold for quartic operators.

\section{Diffusive eigenoperators of the dual Lindbladian quartic in fermions}
\label{sec:states}
We now turn to the determination of the diffusive eigenoperators quartic in fermion, which upon vectorization corresponds to eigenstates with $N_\up=N_\down=2$. We recall that the existence of eigenoperators involving a definite number of fermions is a consequence of the decoupling of the BBGKY hierarchy and is absent in Lindblad equations for which the Hamiltonian part involves interactions, \emph{c.f.} see Ref.~\cite{bauer2017stochastic} for a discussion of hydrodynamic modes in such cases. 
The right eigenstates of the dual Lindbladian can be obtained from the energy eigenstates of the Hubbard model by using the relation \fr{hubbardrel}. In order to obtain a complete set of eigenstates it is necessary to combine the Bethe Ansatz with the SO(4) symmetry of the Hamiltonian \cite{essler1992new,essler1992completeness}. We are interested in describing the late time dynamics, which is governed by eigenstates with eigenvalues close to zero. These were identified in \cite{Medvedyeva2016Exact}, and in the $N_\uparrow=2=N_\downarrow$ sector fall into four classes
\cite{penc2026linear}.

We start by describing the relevant four-particle Bethe states $|\Phi_{R,\bol}\rangle$, which we label by the rapidity variables $\bol=\{\lambda_1,\lambda_2\}$
\begin{align}
|\Phi_{R,\bol}\rangle=\sum_{\boldsymbol{x}}\phi_{R,\bol}(\boldsymbol{x})c^\dagger_{x_1,\downarrow}
c^\dagger_{x_2,\downarrow}c^\dagger_{x_3,\uparrow}c^\dagger_{x_4,\uparrow}|0\rangle\ ,
\end{align}
where the unnormalized wave functions in the sector $x_{Q_1}\leq x_{Q_2}\leq
x_{Q_3}\leq x_{Q_4}$ take the form \cite{essler2005one,deguchi2000thermodynamics,essler1992new}
\begin{align}
\phi_{R,\bol}(\boldsymbol{x})&=\sum_{P\in S_4}{\rm
  sgn}(PQ)\ e^{i\sum_{j=1}^4k_{P_j}x_{Q_j}}\varphi(y_1,y_2|P)\ ,\nn
\varphi(y_1,y_2|P)&=\frac{\lambda_1-\lambda_2-iu}{\lambda_1-\lambda_2}F_P(\lambda_1,y_1)F_P(\lambda_2,y_2)
+\frac{\lambda_2-\lambda_1-iu}{\lambda_2-\lambda_1}F_P(\lambda_2,y_1)F_P(\lambda_1,y_2)\ ,\nn
{F_P(\Lambda,y)}&{=\bigg(\prod_{y'=1}^{y-1}\frac{\sin(k_{P_{y'}})-i\lambda-\frac{u}{2}}{\sin(k_{P_{y'}})-i\lambda+\frac{u}{2}}\bigg)
\frac{1}{\sin(k_{P_y})-i\lambda+\frac{u}{2}}}.
\label{eq:wf}
\end{align}
Here $Q=(Q_1,Q_2,Q_3,Q_4)$ is a permutation of $(1,2,3,4)$,
$u=-\frac{\gamma}{J}$ and $(y_1,y_2)$ are the positions of $(x_1, x_2)$ in $\{x_{Q_1},x_{Q_2},
x_{Q_3},x_{Q_4}\}$. Periodic boundary conditions impose the following quantization conditions, known as "Bethe equations", on the parameters $k_j$ and $\lambda_\ell$
\begin{align}
  &e^{ik_jL}=\frac{i\lambda_1-\sin(k_{j})+\frac{u}{2}}{i\lambda_1-\sin(k_{j})-\frac{u}{2}}
  \frac{i\lambda_2-\sin(k_{j})+\frac{u}{2}}{i\lambda_2-\sin(k_{j})-\frac{u}{2}}\ ,\quad j=1,\dots,4\ ,\nn
  &\prod_{j=1}^4\frac{i\lambda_\ell-\sin(k_{j})+\frac{u}{2}}{i\lambda_{\ell}-\sin(k_{j})-\frac{u}{2}}=
\prod_{k\neq \ell}
\frac{\lambda_\ell-\lambda_k-iu}{\lambda_\ell-\lambda_k+iu}\ ,\quad \ell=1,2.
\label{BAE}
\end{align}
The corresponding eigenvalues of $\mathcal{H}$ and total momentum are
\be
E(\bol)=-2iJ\sum_{n=1}^4\cos(k_n)-4\gamma\ ,\qquad
P(\bol)=\sum_{n=1}^4k_n\ .
\ee

The "hydrodynamic" eigenstates arise from the following Bethe-ansatz configurations. Class I describes scattering states of two $k$-$\Lambda$ strings and gives rise to the hydrodynamic tails of translationally invariant quartic operators. Classes II–IV are SO(4) descendants or special bound-state configurations; although gapless in appropriate limits, their overlaps with the translationally invariant operators considered below vanish.

\subsection{Class I: Scattering states of two \sfix{$k$-$\Lambda$} strings}
\label{sec:scattstates}
The first class we investigate are scattering states of two
$k-\Lambda$ strings of length $1$. These correspond to solutions of \fr{BAE} of the form
\be
\sin k_{1,2}=\Lambda_1\pm\frac{u}{2}\pm\epsilon_1\ ,\quad
\sin k_{3,4}=\Lambda_2\pm\frac{u}{2}\pm\epsilon_3\ ,\quad
\Lambda_{1,2}=i\lambda_{1,2}\ ,\ \lambda_{1,2}\in\mathds{R},
\label{eq:spec_pars}
\ee
where $\epsilon_{1,3}$ are exponentially small in the system size and parametrize the deviations from "ideal strings", \emph{cf.} Refs \cite{essler2005one,Medvedyeva2016Exact}. Substituting \fr{eq:spec_pars} into the expression for the wave functions \ref{eq:wf} we obtain the following results for the most divergent contributions $\varphi(y_1,y_2|P)$ to the wave functions in the different $Q$ sectors:

\noindent \underline{$Q\in\{(1,2,3,4),(2,1,3,4),(2,1,4,3),(1,2,4,3)\}$}
\begin{align}
  \varphi(1,2|2431)&\simeq\frac{\lambda_{12}^2+u^2}{\lambda_{12}^2\epsilon_1\epsilon_3}\simeq \varphi(1,2|2413)
  \ ,\
  \varphi(1,2|4213)\simeq\frac{\lambda_{12}^2+u^2}{\lambda_{12}^2\epsilon_1\epsilon_3}\simeq\varphi(1,2|4231),
\end{align}
where we have defined $ \lambda_{12}=\lambda_1-\lambda_2$.

\noindent\underline{$Q\in\{(1,3,2,4),(1,4,2,3),(2,3,1,4),(2,4,1,3)\}$}
\begin{align}
\varphi(1,3|2143)\simeq&\frac{\lambda_{12}+iu}{\lambda_{12}\epsilon_1\epsilon_3}\ ,\
\varphi(1,3|2413)\simeq\frac{u(i\lambda_{12}-u)}{\lambda_{12}^2\epsilon_1\epsilon_3}\ ,\
\varphi(1,3|4321)\simeq\frac{\lambda_{12}-iu}{\lambda_{12}\epsilon_1\epsilon_3}\ ,\nn
\varphi(1,3|4231)\simeq&-\frac{u(u+i\lambda_{12})}{\lambda_{12}^2\epsilon_1\epsilon_3}\ ,\
\varphi(1,3|2431)\simeq-\frac{u^2+\lambda_{12}^2}{\lambda_{12}^2\epsilon_1\epsilon_3}\simeq
\varphi(1,3|4213)\ .
\end{align}

\noindent\underline{$Q\in\{(1,4,3,2),(1,3,4,2),(2,4,3,1),(2,3,4,1)\}$}
\begin{align}
\varphi(1,4|2143)&\simeq-\frac{\lambda_{12}+iu}{\lambda_{12}\epsilon_1\epsilon_3}\simeq\varphi(1,4|2413)\ ,\
\varphi(1,4|4231)\simeq-\frac{\lambda_{12}-iu}{\lambda_{12}\epsilon_1\epsilon_3}\simeq\varphi(1,4|4321)\ .
\end{align}

\noindent\underline{$Q\in\{(3,1,2,4),(4,1,2,3),(3,2,1,4),(4,2,1,3)\}$}
\begin{align}
\varphi(2,3|2143)&\simeq-\frac{\lambda_{12}+iu}{\lambda_{12}\epsilon_1\epsilon_3}\simeq\varphi(2,3|2413)\ ,\
\varphi(2,3|4231)\simeq-\frac{\lambda_{12}-iu}{\lambda_{12}\epsilon_1\epsilon_3}\simeq\varphi(2,3|4321)\ .
\label{Q4213}
\end{align}


\noindent\underline{$Q\in\{(3,1,4,2),(3,2,4,1),(4,1,3,2),(4,2,3,1)\}$}
\begin{align}
\varphi(2,4|2143)\simeq&\frac{\lambda_{12}+iu}{\lambda_{12}\epsilon_1\epsilon_3}\ ,\
\varphi(2,4|2413)\simeq\frac{u(i\lambda_{12}-u)}{\lambda_{12}^2\epsilon_1\epsilon_3}\ ,\
\varphi(2,4|4321)\simeq\frac{\lambda_{12}-iu}{\lambda_{12}\epsilon_1\epsilon_3}\ ,\nn
\varphi(2,4|4231)\simeq&-\frac{u(u+i\lambda_{12})}{\lambda_{12}^2\epsilon_1\epsilon_3}\ ,\
\varphi(2,4|2431)\simeq-\frac{u^2+\lambda_{12}^2}{\lambda_{12}^2\epsilon_1\epsilon_3}\simeq
\varphi(2,4|4213)\ .
\end{align}

\noindent\underline{$Q\in\{(3,4,1,2),(4,3,1,2),(3,4,2,1),(4,3,2,1)\}$      }
\begin{align}
\varphi(3,4|2413)&\simeq\frac{\lambda^2_{12}+u^2}{\lambda_{12}^2\epsilon_1\epsilon_3}\simeq\varphi(3,4|2431)\ ,\
\varphi(3,4|4231)\simeq\frac{\lambda_{12}^2+u^2}{\lambda_{12}^2\epsilon_1\epsilon_3}\simeq\varphi(3,4|4213)\ .
\end{align}
\subsubsection{Discrete Takahashi equations}
In order to proceed we now neglect the string deviations $\epsilon_{1,2}$ and change variables to
\begin{align}
  k_{1,2}&=\frac{k_+(\lambda_1)}{2}\pm k_-(\lambda_1,J)\ , \quad k_{3,4}=\frac{k_+(\lambda_2)}{2}\pm k_-(\lambda_2,J)\ ,
\end{align}
where
\begin{align}
k_+(\lambda)&=\mathrm{sgn}(\lambda)\ {\rm
  arccos}\big(z_-(\lambda)\big)\ ,\nn
z_-(\lambda)&=\lambda^2+\frac{u^2}{4}-\sqrt{\big(\lambda^2+\frac{u^2}{4}\big)^2+2\big(\lambda^2-\frac{u^2}{4}+\frac{1}{2}\big)}\label{eq:kp}\ ,\nn
k_-(\lambda,J)&={{\rm arcsin}\ \Big(\frac{u}{2\cos\big(\frac{k_+(\lambda)}{2}\big)}\Big).}
\end{align}

One subtlety we have to account for is that the string solution in general only exists for a limited range of momenta. This means that we have to distinguish between the following cases:
\begin{enumerate}
\item{} $\gamma\geq 2J$

Here $k_+(\lambda)$ is a continuous function on
$[-\pi,\pi]$ when $\lambda\in(-\infty,\infty)$.

\item{} $\gamma< 2J$

Here the string solution only exists on part of the Brillouin zone
  \be
  |k_+(\lambda)|\geq 2\ {\rm arccos}(\frac{u^2}{2}-1)\ .
\ee
\end{enumerate}
In the following we will focus on the first case for simplicity. 
In terms of the new variables the Bethe equations read
\be
e^{iLk_+(\lambda_1)}=\frac{\lambda_{12}+iu}{\lambda_{12}-iu}=e^{-iLk_+(\lambda_2)}\ .
\label{eq:Bethekp}
\ee
Importantly $k_+(\lambda)$ is a monotonic function of $\lambda\in\mathds{R}$. Taking the logarithm of \fr{eq:Bethekp} gives
\be
Lk_+(\lambda_a)=\sum_{b\neq
  a}-i\ln\left[-\frac{\lambda_{ab}+iu}{\lambda_{ab}-iu}\right]+2\pi
I_a\ ,
\ee
which allows us to label all solutions $(\lambda_1,\lambda_2)$ by sets of distinct
half-odd integers $(I_1,I_2)$.
The corresponding eigenvalues $E(\lambda_1,\lambda_2)$ of $\mathcal{H}$ are
\be
E(\lambda_1,\lambda_2)=\sum_{a=1}^22\sqrt{\gamma^2-4J^2\cos^2(k_+(\lambda_a)/2)}-4\gamma\ .
\ee

\subsubsection{Hydrodynamic limit of the wave functions}
The eigenstates corresponding to the diffusive eigenoperators occur at $k_+(\lambda_{1,2})\approx\pm\pi$, where we recover a quadratic dispersion
\begin{equation}
    E(\lambda_1,\lambda_2)= -\frac{J^2}{\gamma} \big((|k_+(\lambda_1)|-\pi)^2+(|k_+(\lambda_2)|-\pi)^2\big)+\dots\ .
\end{equation}
We define for $|\lambda_a|\to\infty$
\be
k_+(\lambda_a)=\pi\ \sgn(\lambda_a)+p_a\ ,\quad |p_a|\ll 1\ ,\quad a=1,2.
\ee
In this limit we have
\be
\frac{1}{\lambda_a}=\frac{p_a}{u}+{\cal O}(p_a^3)\ ,\quad e^{ik_-(\lambda_a)}=\frac{p_a}{2u}e^{-i\frac{\pi}{2} \sgn(\lambda_a)}+{\cal O}(p_a^3)\ .
\ee
In order to determine the form of the wave functions in the small-$p_a$ limit we impose 
\be
x_1>x_2\ ,\quad x_3>x_4\ ,
\ee
and construct the wave function for other orderings using the antisymmetry of the wave function. It is useful to define
\be
f_p(x,y)=e^{ip\frac{x+y}{2}}(-1)^y\left(\frac{p}{2u}\right)^{x-y}\ .
\ee
The leading contributions in the various $Q$-sectors are given by

\noindent \underline{$Q=(2,1,4,3)$}
\be
\Phi_{R,\bol}(\boldsymbol{x})\approx\big(1+\frac{u^2}{\lambda_{12}^2}\big)
\left[f_{p_1}(x_3,x_2)f_{p_2}(x_4,x_1)-f_{p_1}(x_4,x_2)f_{p_2}(x_3,x_1)+(p_1\leftrightarrow p_2)
\right].
\label{Phi2143}
\ee

\noindent \underline{$Q=(2,4,1,3)$}
\begin{align}
\Phi_{R,\bol}(\boldsymbol{x})&\approx
\frac{iu}{\lambda_{12}}\big(1+\frac{iu}{\lambda_{12}}\big)f_{p_1}(x_1,x_2)f_{p_2}(x_3,x_4)
-\frac{iu}{\lambda_{12}}\big(1-\frac{iu}{\lambda_{12}}\big)f_{p_1}(x_3,x_4)f_{p_2}(x_1,x_2)\nn
&+\big(1+\frac{u^2}{\lambda_{12}^2}\big)
\left[f_{p_1}(x_3,x_2)f_{p_2}(x_1,x_4)+(p_1\leftrightarrow p_2)\right]\nn
&-\big(1-\frac{iu}{\lambda_{12}}\big)f_{p_1}(x_3,x_1)f_{p_2}(x_4,x_2)
-\big(1+\frac{iu}{\lambda_{12}}\big)f_{p_1}(x_4,x_2)f_{p_2}(x_3,x_1).
\label{Phi2413}
\end{align}

\noindent \underline{$Q=(2,4,3,1)$}
\begin{align}
\Phi_{R,\bol}(\boldsymbol{x})\approx&-\big(1+\frac{iu}{\lambda_{12}}\big)
\left[f_{p_1}(x_4,x_2)f_{p_2}(x_1,x_3)-f_{p_1}(x_3,x_2)f_{p_2}(x_1,x_4)\right]\nn
&-\big(1-\frac{iu}{\lambda_{12}}\big)
\left[f_{p_1}(x_1,x_3)f_{p_2}(x_4,x_2)-f_{p_1}(x_1,x_4)f_{p_2}(x_3,x_2)\right].
\label{Phi2431}
\end{align}

\noindent \underline{$Q=(4,2,1,3)$}
\begin{align}
\Phi_{R,\bol}(\boldsymbol{x})\approx&\big(1+\frac{iu}{\lambda_{12}}\big)
\left[f_{p_1}(x_1,x_4)f_{p_2}(x_3,x_2)-f_{p_1}(x_2,x_4)f_{p_2}(x_3,x_1)\right]\nn
&+\big(1-\frac{iu}{\lambda_{12}}\big)
\left[f_{p_1}(x_3,x_2)f_{p_2}(x_1,x_4)-f_{p_1}(x_3,x_1)f_{p_2}(x_2,x_4)\right].
\end{align}

\noindent \underline{$Q=(4,2,3,1)$}
\begin{align}
\Phi_{R,\bol}(\boldsymbol{x})&\approx
-\frac{iu}{\lambda_{12}}\big(1-\frac{iu}{\lambda_{12}}\big)f_{p_1}(x_1,x_2)f_{p_2}(x_3,x_4)
+\frac{iu}{\lambda_{12}}\big(1+\frac{iu}{\lambda_{12}}\big)f_{p_1}(x_3,x_4)f_{p_2}(x_1,x_2)\nn
&+\big(1+\frac{u^2}{\lambda_{12}^2}\big)
\left[f_{p_1}(x_3,x_2)f_{p_2}(x_1,x_4)+(p_1\leftrightarrow p_2)\right]\nn
&-\big(1-\frac{iu}{\lambda_{12}}\big)f_{p_1}(x_1,x_3)f_{p_2}(x_2,x_4)
-\big(1+\frac{iu}{\lambda_{12}}\big)f_{p_1}(x_2,x_4)f_{p_2}(x_1,x_3)\ .
\label{Phi4231}
\end{align}

\noindent \underline{$Q=(4,3,2,1)$}
\begin{align}
\Phi_{R,\bol}(\boldsymbol{x})&\approx\big(1+\frac{u^2}{\lambda_{12}^2}\big)
\left[f_{p_1}(x_1,x_4)f_{p_2}(x_2,x_3)-f_{p_1}(x_2,x_4)f_{p_2}(x_1,x_3)+(p_1\leftrightarrow p_2)\right].
\label{Phi4321}
\end{align}
Imposing the total momentum to be zero corresponds to $p=p_1=-p_2$. The above expressions then simplify
\be
f_p(x_1,y_1)f_{-p}(x_2,y_2)=\big(\frac{p}{2u}\big)^{x_1-y_1+x_2-y_2}(-1)^{y_1+x_2}e^{ip\frac{x_1-x_2+y_1-y_2}{2}} .
p\ee
The results for the zero momentum sector (and $x_1>x_2$, $x_3>x_4$) can be written as
\begin{align}
\phi_{R,\{\lambda,-\lambda\}}(\boldsymbol{x})\simeq-2\kappa(\boldsymbol{x})(-1)^{x_{Q_2}+x_{Q_3}} \left(\frac{J}{2\gamma}\right)^{|x_1-x_3|+|x_2-x_4|} p^{\xi(\boldsymbol{x})}\ ,
\label{phiRasy}
\end{align}
where we defined $h(x)=1-\mathrm{mod}(x,2)$ and
\begin{align}
\xi(\boldsymbol{x}) &= \begin{cases}
        |x_1-x_3|+|x_2-x_4| + h(x_1-x_2)+h(x_3-x_4)& x_2\geq x_3\ \mathrm{or}\ x_4\geq x_1\ , \\
         |x_1-x_3|+|x_2-x_4| + \mathrm{mod}(x_1-x_2+x_3-x_4,2)& \mathrm{else}\ ,
    \end{cases}\nn
    \kappa(\boldsymbol{x}) &= \begin{cases}
        2\left(\frac{(x_1-x_2)}{2 i}\right)^{h(x_1-x_2)}\left(\frac{(x_3-x_4)}{2 i}\right)^{h(x_3-x_4)} & x_2\geq x_3\ \mathrm{or}\ x_4\geq x_1 \ ,\nn
        \left(\frac{x_1-x_2+x_3-x_4-1}{2i}\right)^{\mathrm{mod}(x_1-x_2+x_3-x_4,2)}& \mathrm{else}\ .
        \end{cases}
\end{align}
We note that $\xi(\boldsymbol{x})$ is always an even integer.

\subsubsection{Left eigenstates and normalisation}
The Hamiltonian is non-Hermitian, and obeys the relation
\begin{equation}
    \mathcal{H}(J)^\dagger = \mathcal{H}(-J)\ .
\end{equation}
For the eigenstates of interest here the eigenvalues are real $E^*(\bol)=E(\bol)$ and independent of the sign of $J$. The Hermitian conjugate of the right eigenvalue equation reads
\be
\langle\Phi_{R,\bol}|{\cal   H}(-J)=E(\bol)\langle\Phi_{R,\bol}|\ .
\ee
This implies that 
\be
\langle\Phi_{L,\bol}|=\langle\Phi_{R,\bol}|\Big|_{J\rightarrow -J}.
\ee
Left and right eigenstates form an orthogonal set \fr{LR}, where the normalization can be computed using
Ref.~\cite{gohmann1999hubbard}
\be
|A_{\bol}|^2\equiv\epsilon_1^2\epsilon_3^2|\mathcal{A}_{\bol}|^2
\simeq\frac{\epsilon_1^2\epsilon_3^2{|u^2|}}{|16L^2 (\cos k_1+\cos k_2)(\cos
  k_3+\cos
  k_4)\big(1+\frac{u^2}{\lambda_{12}^2}\big)|}\ .
\ee

\subsection{Class II: \sfix{$k$-$\Lambda$} strings of length 2}
\label{sec:boundstates}
The second class of eigenstates that has eigenvalues of $\mathcal{H}$ close to zero corresponds to four-particle bound states of the form
\begin{align}
\Lambda_1&=i\lambda+\frac{u}{2}+\delta\ ,\ \Lambda_2=i\lambda-\frac{u}{2}-\delta\ ,\nn
\sin k_1&=i\lambda +u+\epsilon_1\ ,\
\sin k_2=i\lambda +\epsilon_2\ ,\
\sin k_3=i\lambda -\epsilon_2\ ,\
\sin k_4=i\lambda -u-\epsilon_1\ .
\end{align}
We find that the most singular part of the wave function for large $L$ is given by
\begin{equation}
    \phi_{R,\bol}(\boldsymbol{x})=\begin{cases}
        \alpha\ \sgn(Q)(-1)^{y_1+y_2} (e^{i\sum_{j=1}^4k_{R_j} x_{Q_j}}-R\rightarrow\bar{R}) & \mathrm{if}\ (y_1,y_2)\neq(1,2),(3,4)\ , \\
        0 & \mathrm{else}\ ,
    \end{cases}
    \label{eq:wf4bound}
\end{equation}
where $\alpha=4\delta/[(\delta^2-\eps_2^2)(\delta-\eps_1)]$ and we have defined $R = (4,2,3,1)$,  $\bar{R}= (4,3,2,1)$. 
Dropping the exponentially small finite-size corrections we have
\begin{align}
k_1&=\begin{cases}
{\rm arcsin}(i\lambda +u) & \text{ if } \lambda>0\ ,\\
\pi-{\rm arcsin}(i\lambda +u)& \text{ if } \lambda<0\ ,
\end{cases}\ ,\quad
k_4=\begin{cases}
\pi-{\rm arcsin}(i\lambda -u) & \text{ if } \lambda>0\ ,\\
{\rm arcsin}(i\lambda -u)& \text{ if } \lambda<0\ ,
\end{cases}\nn
k_2&=\pi-k_3=\begin{cases}
{\rm arcsin}(i\lambda)& \text{ if } \lambda>0\ ,\\
\pi-{\rm arcsin}(i\lambda)& \text{ if } \lambda<0\ .
\end{cases}
\end{align}
We have
\be
{\rm Im}(k_1)=-{\rm Im}(k_4)>
{\rm Im}(k_2)=-{\rm Im}(k_3)\geq 0\ ,
\ee
which ensures that the wave functions \fr{eq:wf4bound} decays exponentially in $x_{Q_4}-x_{Q_1}$ and describes a four-particle bound state. The total momentum and Lindbladian eigenvalue are
\begin{align}
P&=-2\ \sgn (\lambda)\ {\rm Im}\big({\rm arcsinh}(|\lambda| -iu)\big)\ ,\nn
E(P)&=4\sqrt{\gamma^2-J^2\sin^2(P/2)}-4\gamma\ .
\end{align}
The normalization of \fr{eq:wf4bound} is found to be
\be
|A_{\bol}|^2\equiv\frac{|\mathcal{A}_{\lambda}|^2}{|\alpha|^2}
\simeq\frac{1}{|\alpha|^2}\frac{|u^3|}{|8L\cos (k_1)\cos (k_2)\cos(
  k_3)\cos(
  k_4)\sum_{j=1}^4\frac{1}{\cos(k_j)}|}\ .
\ee

\subsection{Class III: \sfix{$\eta$}-pairing descendants of eigenstates with \sfix{$N_\uparrow=1=N_\downarrow$}}

The third class of eigenstates are $\eta$-pairing descendants of Bethe states in the two-particle sector. For large $L$ we have \cite{penc2026linear}
\begin{align}
\frac{1}{\sqrt{L-2}}\eta^\dagger |\Phi_{R,\lambda}\rangle&=\sum_{x_1,x_2,x_3}
\frac{(-1)^{x_2} }{\sqrt{L(L-2) \tanh(\eta(P))}}
 \chi_P(x_1,x_3)\ c^\dagger_{x_1,\downarrow} c^\dagger_{x_2,\downarrow} c^\dagger_{x_2,\up} c^\dagger_{x_3,\up}|0\rangle\ ,\nn
\chi_{P}(x,y)&=e^{i P\frac{x+y}{2}}\Big[ e^{(i\kappa(P)-\eta(P))|x-y|}+ e^{iLP/2} e^{(i\kappa(P)-\eta(P))(L-|x-y|)}\Big]\ ,\nn
\kappa(P)&= -\frac{\pi}{2}\sgn\left(J\cos(\frac{P}{2})\right)\ ,\quad\eta(P)=\mathrm{arccosh}\Big(\big|\frac{\gamma}{2J\cos(P/2)}\big|\Big)\ ,
\end{align}
where $P$ fulfils $\exp(iPL)=1$. These states have eigenvalues
\begin{equation}
E(P) = 2\sqrt{\gamma^2 - 4J^2 \cos^2(P/2)}-2\gamma\ .
\end{equation}

\subsection{Class IV: \sfix{$\eta$}-pairing descendant of the fermion vacuum}

The final quartic eigenoperator of interest corresponds to the $\eta$-pairing descendant of the vacuum state
\begin{equation}
|N^2\rangle=\frac{1}{\sqrt{2L(L-1)}}    (\eta^+)^2 |0\rangle
\label{eta2}
\end{equation}
As the vacuum state has zero eigenvalue and the $\eta$-pairing is a symmetry of the Hamiltonian we have
\begin{equation}
    E = 0\ .
\end{equation}
Undoing the vectorization we find that \fr{eta2} corresponds to the eigenoperator
\be
\hat{N}^2=\frac{1}{\sqrt{2L(L-1)}}(\sum_{j=1}^Ln_j)^2\ .
\ee

\section{Hydrodynamic interactions}
\label{sec:eff_theory}
In Section \ref{sec:states} we have derived explicit expressions for the hydrodynamic eigenoperators quartic in fermions. We will now show how these are related in the small momentum limit to the hydrodynamic eigenoperators quadratic in fermions previously derived in Ref.~\cite{penc2026linear}. In the large-$L$ limit the latter reads
\begin{equation}
\hat{\Phi}^{(2)}_{\pi+p}= \sum_{x_1,x_2} (-1)^{\min(0,x_1-x_2)} e^{i p \frac{x_1+x_2}{2}}(\frac{p}{2u})^{|x_1-x_2|} c^\dagger_{x_1}c_{x_2}\ .
\end{equation}
As the amplitudes decay exponentially with respect to the distance $|x_1-x_2|$ it is useful to introduce operators $\hat{\phi}_{\pi+p}(X)$ that are concentrated around the centre of mass $X=\frac{x_1+x_2}{2}$ by
\begin{equation}
\hat{\Phi}^{(2)}_{\pi+p}= \sum_{X\in \mathbb{Z}/2}e^{i p X} \widehat{\phi}^{(2)}_{\pi+p}(X)\ .
\end{equation}
For small $p$ we have
\be
\mathcal{L}^*[\hat{\Phi}^{(2)}_{\pi+p}]\simeq -\mathcal{D}p^2\hat{\Phi}^{(2)}_{\pi+p}\ ,
\ee
where $\mathcal{D} = J^2/\gamma $ is the diffusion constant. In section \fr{sec:scattstates} we have determined eigenoperators quartic in fermions: they are described in terms of scattering states of two $k$-$\Lambda$ strings and are characterized by two momentum variables $p_{1,2}=k_+(\lambda_{1,2})-\pi\sgn(k_+(\lambda_{1,2}))$. For small $p_{1,2}$ we have 
\be
\mathcal{L}^*[\hat{\Phi}^{(4)}_{p_1,p_2}]\simeq -\mathcal{D}(p_1^2+p_2^2)\hat{\Phi}^{(4)}_{p_1,p_2}\ .
\ee
The position representation of $\hat{\Phi}^{(4)}_{p_1,p_2}$ is
\be
\hat{\Phi}^{(4)}_{p_1,p_2}=\sum_{\boldsymbol{x}}(-1)^{x_1+x_2}\Phi_{R,\{\lambda_1(p_1),\lambda_2(p_2)\}}(\boldsymbol{x})\ c^\dagger_{x_3}c^\dagger_{x_4}c_{x_1}c_{x_2}\ ,
\label{Phi4full}
\ee
where $\lambda(p)$ is obtained by inverting the equation $\pi\ \sgn(\lambda)+p=k_+(\lambda)$.
For small momenta $|p_{1,2}|\ll 1$ it follows from \fr{Phi2143}-\fr{Phi4321} that these can be approximately written as
\be
\hat{\Phi}^{(4)}_{p_1,p_2} \approx \sum_{X>Y} \left[e^{i (p_1 X+p_2 Y)}
\left(1-i\frac{p_1p_2}{p_2-p_1}\right)\widehat{\phi}_{\pi+p_1}(X) \widehat{\phi}_{\pi+p_2}(Y)+(p_1\leftrightarrow p_2)\right],
\label{Phi4Phi2}
\ee
where $X,Y\in\mathbb{Z}/2$ are the centre-of-mass coordinates of the two two-particle bound states. The wave function in \fr{Phi4Phi2} exactly reproduces the small momentum limit of the leading contribution to $\Phi_{R,\bol} (\boldsymbol{x})$ in \fr{Phi4full} when the two bound states are well-separated, but disagrees at cubic order ${\cal O}(p_{1,2}^3)$ when they overlap, \emph{cf.} \fr{Phi2143}-\fr{Phi4321}. If we think about the 
quadratic eigenoperator $\hat{\Phi}^{(2)}_{\pi+p}$ as the "elementary" diffusive mode, then \fr{Phi4Phi2} implies that there is an interaction between diffusive modes, characterized by the scattering phase
\be
S(p_1,p_2)=\frac{p_2-p_1+ip_1p_2}{p_2-p_1-ip_1p_2}\ .
\ee
This is as expected in agreement with the discrete Takahashi equations \fr{eq:Bethekp} (i.e. the Bethe equations for $k$-$\Lambda$ strings).

\section{Hydrodynamic projections of translationally invariant operators}
Having determined the diffusive eigenoperators in the four-particle sector we are now in a position to work out the late-time structure of neutral Heisenberg-picture operators quartic in fermions. For the sake of definiteness we consider the particular operators
\be
\hat{\Psi}^{(Q)}_{\{d,d,0\}}=\sum_{x}n_xn_{x+d}e^{iQx}\ ,
\ee
where we will take the limit $Q\to 0$ in the end. The vectorized form is
\be
|\Psi^{(Q)}_{\{d,d,0\}}(t)\rangle=\sum_{\boldsymbol{x}}{\Psi}^{(Q)}_{\d0}(\boldsymbol{x},t)c^\dagger_{x_1,\up}c^\dagger_{x_2,\up}
c^\dagger_{x_3,\down}
c^\dagger_{x_4,\down}|0\rangle
\label{wavefnt}
\ee
where the initial wave function takes the simple form
\be
{\Psi}^{(Q)}_{\d0}(\boldsymbol{x},t=0)=(-1)^{d}\delta_{x_2,x_1+d}\delta_{x_3,x_1+d}\delta_{x_4,x_1}\ e^{iQx_1}\ .
\ee
The overlaps with eigenstates of $\mathcal{H}$ are 
\be
\langle\Phi_{L,\alpha}|\Psi_{\d0}\rangle=-4(-1)^d\sum_x e^{iQx}\Phi^*_{L,\alpha}(x,x+d,x,x+d)\ .
\ee
The contributions of eigenstates in classes II-IV to the time evolution \fr{specrep} of translationally invariant operators all vanish and the late-time hydrodynamic tails are entirely determined by eigenstates in class I. The overlaps with these are given by
\begin{align}
\langle\Phi_{L,\bol}|\Psi^{(Q)}_{\d0}\rangle&=4(-1)^dL\delta_{\nu_1+\nu_2,Q}e^{-iQd/2}\Big[
    \frac{\lambda_{12}+iu}{\lambda_{12}}e^{-i(\nu_1-\nu_2)d/2}+\frac{\lambda_{12}-iu}{\lambda_{12}}e^{i(\nu_1-\nu_2)d/2}\nn
    &\qquad-2e^{-id\big(k^*_-(\lambda_1,-J)+k^*_-(\lambda_2,-J)\big)}\Big]e^{iQx}\equiv
L\delta_{\nu_1+\nu_2,Q} \psi_0(\bol,d;Q)\ ,
\end{align}
where we have defined $\nu_a=k_+(\lambda_a)$.
The contribution from class I states to $|\Psi^{(Q)}_{\d0}(t)\rangle$ is then
\begin{align}
|\Psi^{(Q)}_{\d0}(\gamma t\gg 1)\rangle&=\sum_{I_1<I_2}|A_{\bol}|^2\langle\Phi_{L,\bol}|\Psi^{(Q)}_{\d0}\rangle
e^{E(\bol)t}|\Phi_{R,\bol}\rangle\nn
&=\sum_{I_1<I_2}L\delta_{\nu_1+\nu_2,Q}|A_{\bol}|^2\psi_0(\bol,d;Q)
e^{E(\bol)t}\Phi_{R,\bol}(\boldsymbol{x}) c^\dagger_{x_1,\downarrow}c^\dagger_{x_2,\downarrow}c^\dagger_{x_3,\uparrow}c^\dagger_{x_4,\uparrow}|0\rangle\ .
\end{align}

\subsection{Bethe equations and large-\sfix{$L$} limit}
The Bethe equations read
\begin{align}
\frac{2\pi
  I_1}{L}&=k_+(\lambda_1)+\frac{i}{L}\ln\left[\frac{iu+\lambda_{12}}{iu-\lambda_{12}}\right]\ ,\nn
\frac{2\pi
  I_2}{L}&=k_+(\lambda_2)-\frac{i}{L}\ln\left[\frac{iu+\lambda_{12}}{iu-\lambda_{12}}\right]\ .
  \label{eq:bethelargeL}
\end{align}
Here the ``interaction term'' is a small correction when $I_a/L$ is
order $L^0$. Assuming that we therefore can neglect it when turning
sums into integrals for large $L$ we have
\be
\frac{1}{L^2}\sum_{I_1<I_2}\longrightarrow\int_{-\pi}^{\pi}\frac{d\nu_1
  d\nu_2}{2(2\pi)^2}\ .
\ee
We recall that we have assumed for simplicity that $\gamma> 2J$, so that
$-\pi<\nu\leq\pi$, i.e. the 2-particle bound state exists in the entire
Brillouin zone. 
Similarly we obtain
\begin{align}
L\delta_{\nu_1+\nu_2,Q}&\longrightarrow
2\pi\ \delta(\nu_1+\nu_2-Q)\ ,\nn
\lambda_a&\longrightarrow
\lambda(\nu_a)=\mathrm{sgn}(\nu_a )\sqrt{\big(\cos^2\frac{\nu_a}{2}-1\big)\big(1-\frac{u^2}{4\cos^2\frac{\nu_a}{2}}\big)}\ ,\nn
|A_{\bol}|^2&\longrightarrow \frac{1}{L^2}|A_{\boldsymbol{\nu}}|^2=
\frac{1}{L^2\Big(1+\big(\frac{u}{\lambda(\nu_1)-\lambda(\nu_2)}\big)^2\Big)}\prod_{j=1}^2 \frac{\gamma}{4\sqrt{\gamma^2-4J^2\cos^2\frac{\nu_j}{2}}}\ .
\end{align}
\subsection{Integral representation and late-time limit}
Turning sums into integrals we arrive at
\begin{align}
|\Psi_{\d0}^{(Q)}(t)\rangle&=\int_{-\pi}^{\pi}\frac{d\nu_1 d\nu_2}{4\pi}\delta(\nu_1+\nu_2-Q)|A_{\bon}|^2
\psi_0(\bol,d;Q)
e^{E(\bol)t}\nn
&\qquad\qquad\times\ \sum_{\boldsymbol{x}}\Phi_{R,\bol}(\boldsymbol{x}) \ c^\dagger_{x_1,\downarrow}c^\dagger_{x_2,\downarrow}c^\dagger_{x_3,\uparrow}c^\dagger_{x_4,\uparrow}|0\rangle\ ,
\end{align}
where
\be
\bol=\big\{\lambda(\nu_1),\lambda(\nu_2)\big\}.
\ee
The most interesting case is when $Q=0$, as the eigenvalue is gapped otherwise, and the integral decays exponentially in time. We can now use the delta-function to carry out the integral over
$\nu_2$, leaving us with an integral representation involving a single
integral. The delta-function imposes $\nu_2=-\nu_1$ for $Q=0$, which
gives
\begin{align}
\Psi_0(\boldsymbol{\lambda},d;Q\to 0)&\simeq
8(-1)^d\big[\cos(\nu_1
  d)-e^{-id(k^*_-(\lambda_1,-J)+k^*_-(-\lambda_1,-J))}-\frac{u\sin(\nu_1 d)}{2\lambda(\nu_1)}\Big]\equiv \chi_1(\nu_1) ,\nn
|A_{\boldsymbol{\nu}}|^2&\simeq
\frac{\gamma^2}{16(\gamma^2-4J^2\cos^2\frac{\nu_1}{2})\Big(1+\big(\frac{u}{2\lambda(\nu_1)}\big)^2\Big)}\equiv |A_{\nu_1}|^2\ ,
\end{align}
In order to proceed it is useful to focus on the amplitudes \fr{wavefnt}
\be
\Psi_{\d0}^{(0)}(\boldsymbol{x},t)\simeq \int_{-\pi}^{\pi}\frac{d\nu_1 }{\pi}
\chi_1(\nu_1)\Phi_{R,\bol}(\boldsymbol{x})|A_{\nu_1}|^2
e^{E(\nu_1) t}\ ,
\ee
where without loss of generality we have imposed $x_1>x_2,x_3>x_4$ and 
$\Phi_{R,\bol}(\boldsymbol{x})$ are given by \fr{phiRasy}.
The integral can be approximated at late times by expanding the integrand around $\pm\pi$ using $\nu_1=\pm\pi+p$ and then taking the integration boundaries for the resulting $p$-integral to $\pm \infty$. The expansion gives
\begin{align}
    |A_\nu|^2 \approx \frac{1}{16}, \quad \chi_1(\nu_1) \approx 8,\quad E({\nu_1}) \approx -\frac{2J^2}{\gamma}(\nu_1\mp\pi)^2\ .
\end{align}
These can be substituted back to the integral which can be evaluated 
\begin{align}
\Psi_{\d0}^{(0)}(\boldsymbol{x},t)&\simeq \left[-2\kappa(\boldsymbol{x}){(-1)^{x_{Q_2}+x_{Q_3}}}\big(\frac{J}{2\gamma}\big)^{|x_1-x_3|+|x_2-x_4|}\right]\int_{-\infty}^{\infty}\frac{dp}{2\pi}
e^{-\frac{2J^2t}{\gamma}p^2}p^{\xi(\boldsymbol{x})}\nn
&=\left[-2\kappa(\boldsymbol{x}){(-1)^{x_{Q_2}+x_{Q_3}}}\big(\frac{J}{2\gamma}\big)^{|x_1-x_3|+|x_2-x_4|}\right]\frac{\Gamma\big(\frac{1+\xi(\boldsymbol{x})}{2}\big)}{2\pi}
\left[\frac{\gamma}{2J^2t}\right]^{\frac{1+\xi(\boldsymbol{x})}{2}}\ .
\label{tail}
\end{align}
Eq. \fr{tail} gives the leading asymptotic behaviour in $t$ for a given, fixed $\boldsymbol{x}$ in the limit $L\to\infty$ and is the main result of this work. The slowest decay corresponds to $\xi(\boldsymbol{x})=0$. Isolating this contribution under undoing the vectorization gives
\begin{equation}
\Big(\sum_{x=1}^Ln_xn_{x+d}\Big)(t)= \sum_{L\gg x\neq y}\sqrt{\frac{\gamma}{2\pi J^2t}}\ n_xn_{y} + \dots
\label{eq:slowest_decay}
\end{equation}
Here the restriction on the sum on the right hand side arises from the way we have performed the asymptotic expansion, i.e. taking $L\to\infty$ at fixed $x$ and $y$. Eqn \fr{eq:slowest_decay} is a key result of this work.

Other translationally invariant operators $\hat{\Psi}_{\boldsymbol{d}}$ defined in \fr{psid} can be analyzed in the same way. Like in the case of local operators quadratic in fermions one obtains power-law tails with exponents that depend on the details of the operator considered, i.e. on the vector $\boldsymbol{d}$. The physical reason for this behaviour is the small size of the bound states that form the quartic eigenoperator of the dual Lindbladian, \emph{cf.} the discussion in section \ref{sec:tails}.

\section{Diffusive continuum scaling limit}
The late time limit we have considered above applies to the lattice model itself and is very different from continuum scaling limits that have been considered for the QSSEP and similar models \cite{bernard2022dynamics,hruza2023coherent,bernard2026quantum,albert2026universal}. For the tight-binding model with dephasing noise this is achieved by introducing a lattice spacing $a_0$ and considering the continuum limit
\begin{align}
a_0&\to 0\ ,\quad x=ja_0\text{ fixed}\ ,\mathcal{L}=La_0\text{ fixed}\ ,\nn
\gamma,J&\to \infty\ ,\quad \frac{J^2a_0^2}{\gamma}=\mathcal{D}\text{ fixed}.
\label{contscaling}
\end{align}
In this limit the diffusive bound state eigenmode becomes completely local in space, so that the only operators with non-vanishing hydrodynamic projections are of the form $n_{x_1}\dots n_{x_k}$. The diffusion constant and their equations of motion can be understood by employing a Schrieffer-Wolff similarity transformation \cite{macdonald1988t,kessler2012generalized} that projects the non-Hermitian Hamiltonian \fr{HHub} on the subspace with only doubly occupied sites. This is because the scaling limit \fr{contscaling} is compatible with $J/\gamma\rightarrow 0$. 
It is convenient to introduce Hubbard operators
\be
X_j^{ab}=|a\rangle_j {}_j\langle a|\ ,\quad |\sigma\rangle_j=c^\dagger_{j,\sigma}|0\rangle_j\ ,\ \sigma\in\{\up,\down\}\ ,\quad|2\rangle_j=c^\dagger_{j,\up}c^\dagger_{j,\down}|0\rangle_j\ .
\ee
In the large-$\gamma$ limit we then have
\begin{align}
e^{S}{\cal H}e^{-S}=&2\gamma\sum_j\big[(n_{j,\up}-\frac{1}{2})(n_{j,\down}-\frac{1}{2})
-\frac{1}{4}\big]-iJ\sum_{j,\sigma}\big(X_j^{\sigma 0}X_{j+1}^{0\sigma}-
X_j^{\sigma 2}X_{j+1}^{2\sigma}+{\rm h.c.}\big)\nn
&-\frac{J^2}{\gamma}\sum_j\big( X_j^{02}X_{j+1}^{20}+X_j^{20}X_{j+1}^{02}+X_j^{00}X_{j+1}^{22}+X_j^{22}X_{j+1}^{00}\big)+\dots.
\end{align}
The first term projects onto configurations involving only empty and double occupied sites, the third term describes a symmetric simple exclusion process, and the higher order terms will not contribute in the scaling limit. 
This implies that in order to describe the diffusive scaling limit we can consider time evolution on the space with only doubly occupied and empty sites with
\be
\mathcal{H}'=-\frac{J^2}{\gamma}\sum_j\big( X_j^{02}X_{j+1}^{20}+X_j^{20}X_{j+1}^{02}+X_j^{00}X_{j+1}^{22}+X_j^{22}X_{j+1}^{00}\big)\ .
\ee
Undoing the vectorization this gives an equation of motion
\be
\frac{d}{dt}n_j=\frac{J^2}{\gamma}(n_{j+1}+n_{j-1}-2n_j)+o(J^2/\gamma)\ .
\ee
Finally we can take the continuum scaling limit by defining
\be
c_j\rightarrow \sqrt{a_0}c(x)\ ,\quad x=ja_0.
\label{NCL}
\ee
This gives a diffusion equation for $n(x,t)=c^\dagger(x,t)c(x,t)$
\be
\partial_t n(x,t)=\mathcal{D}\partial_x^2n(x,t)\ .
\ee

\section{Conclusions}
In this work we have continued our analysis \cite{penc2026linear} of the Heisenberg-picture dynamics in the tight-binding model with dephasing noise. We used the Yang-Baxter integrability of the model to explicitly construct diffusive eigenoperators quartic in fermions and show that they can be understood in terms of diffusive eigenoperators quadratic in fermions subject to interactions. Using these results we derived explicit closed-form expressions for the late-time behaviour of charge-neutral Heisenberg-picture operators quartic in fermions. We showed that these operators in general have non-vanishing hydrodynamic tails, proving a conjecture put forward in \cite{penc2026linear}.
The structure of the hydrodynamic tails for operators quadratic and quartic in fermions is surprisingly complex and defies expectations based on symmetry arguments. We have provided a physical explanation for this fact in terms
of the spatial sizes of the hydrodynamic eigenoperators of the Lindbladian.
These depend on the momentum but importantly are comparable to the lattice spacing. This results in strong differences in hydrodynamic projections of microscopic operators. In contrast, bounds imposed by symmetry constraints are very loose. These findings apply equally to hydrodynamic tails in the other models with decoupled BBGKY hierarchies investigated in \cite{penc2026linear}.
We conjecture that this is a much more general principle that applies to other lattice models exhibiting diffusive dynamics.

Our work opens several avenues for further investigation. First, one should investigate whether there are examples of diffusive Lindblad equations or dissipative quantum circuits with decoupled BBGKY hierarchies that exhibit hydrodynamic eigenoperators bilinear in fermions whose sizes $\ell$ are large compared to the lattice spacing. In such cases one might expect that the structure of the hydrodynamic tails for fermion bilinears with supports that are small compared to $\ell$ allow for a description in terms of the naive continuum limit \fr{NCL} combined with symmetry arguments.
Second, it would be interesting to investigate the size of hydrodynamic eigenmodes in diffusive Floquet circuits such as \cite{matthies2026thermalization}, and its implications on hydrodynamic projections. Work along these lines is under way \cite{inprep}. 
\section*{Acknowledgements}
This work was supported in part by the EPSRC under grant EP/X030881/1. We are grateful to Curt von Keyserlingk for stimulating discussions.


\bibliography{bibliography}
\end{document}